\documentclass[showpacs,amsmath,amssymb,aps,showkeys,floatfix,a4paper]{revtex4}

\usepackage[dvips]{graphicx}
\usepackage{dcolumn}
\usepackage{bm}
\usepackage{epsfig}
\usepackage{amsfonts}
\usepackage{amssymb,amscd}

\def\lsim{\raise0.3ex\hbox{$<$\kern-0.75em\raise-1.1ex\hbox{$\sim$}}}
\def\gsim{\raise0.3ex\hbox{$>$\kern-0.75em\raise-1.1ex\hbox{$\sim$}}}

\def\odd{{I\!\!\!O}}
\def\beq{\begin{equation}}
\def\eeq{\end{equation}}
\def\bea{\begin{eqnarray}}
\def\eea{\end{eqnarray}}
\def\bq{\begin{quote}}
\def\eq{\end{quote}}

\def\gappeq{\mathrel{\rlap {\raise.5ex\hbox{$>$}}
{\lower.5ex\hbox{$\sim$}}}}

\def\lappeq{\mathrel{\rlap{\raise.5ex\hbox{$<$}}
{\lower.5ex\hbox{$\sim$}}}}

\def\Toprel#1\over#2{\mathrel{\mathop{#2}\limits^{#1}}}

\newcommand{\rb}{\mbox{\boldmath $b$}}

\begin{document}


\title{Searching for the  Odderon  in the diffractive $f_2(1270)$ photoproduction at $pA$ collisions}

\author{V.~P. Gon\c{c}alves}
\email{barros@ufpel.edu.br}
\affiliation{High and Medium Energy Group, \\
Instituto de F\'{\i}sica e Matem\'atica, Universidade Federal de Pelotas\\
Caixa Postal 354, CEP 96010-900, Pelotas, RS, Brazil}
\date{\today}

\begin{abstract}
Although the  Odderon is an unambiguous prediction of Quantum Chromodynamics, its existence  still was not  confirmed experimentally.  One of the processes where the Odderon contribution is expected to be dominant is the diffractive photoproduction of  tensor mesons. In this paper we study the diffractive $f_2(1270)$ photoproduction  in $pA$ collisions at LHC and RHIC considering the model of the stochastic vacuum to treat the non - perturbative process. We demonstrate that this process dominates the $f_2$ production at mid - rapidities, and  large values for the cross section are predicted. Our results indicate that the experimental analysis of this final state is, in principle, feasible and can be used to probe the Odderon.
\end{abstract}

\pacs{12.38.Aw, 13.85.Lg, 13.85.Ni}
\keywords{Quantum Chromodynamics, Meson production,  Photon - induced interactions}

\maketitle


The recent TOTEM data \cite{totem} for the parameter $\rho$, which describes the ratio  between the real and imaginary parts of the scattering amplitude, has motivated an intense debate about the possible contribution of the Odderon \cite{
Martynov:2017zjz, Khoze:2018bus,Broilo:2018els, Troshin:2018ihb,Broniowski:2018xbg,Broilo:2018qqs,Gotsman:2018buo,Csorgo:2018uyp,Goncalves:2018nsp}. Such object is a natural prediction of the Quantum Chromodynamics (QCD), has a $C$-odd parity  and determines the hadronic cross section difference between the direct and crossed channel processes at very high energies (For a review see Ref. \cite{ewerz}). In particular, the Odderon contribution is expected to be important in the dip region  of the differential elastic cross section  
\cite{dosch_ewerz,Ster:2015esa} as well for the description of the diffractive photoproduction of pseudoscalar mesons 
\cite{Kilian:1997ew,nac,nac_odd,ckms,bbcv,bbcv2,vic_odderon1,vic_odderon2,vicbru_etac} (See also Ref. \cite{pheno_odderon} for other possible probes of the Odderon). The last expectation can be easily understood: as the real photon has negative $C$ parity,  its transformation into a diffractive final state system of positive $C$ parity  requires the $t$-channel exchange of an object of negative $C$ parity. It implies that a Pomeron exchange, which carries $C$-even parity, cannot contribute to this process. Therefore, it can only be mediated by the exchange of an Odderon. 
In particular,  the diffractive $\eta_c$ photoproduction have been studied in the last years considering $ep$ collisions  at HERA \cite{ckms,bbcv} and $pp/pA/AA$ collisions at the LHC \cite{vic_odderon1,vic_odderon2} and assuming that  
Odderon is a $C$-odd compound state of three reggeized gluons described by  the Bartels - Kwiecinski - Praszalowicz (BKP) equation \cite{bkp}, which resums terms of the order $\alpha_s(\alpha_s \log s)^n$ with arbitrary $n$ in which three gluons in a $C = -1$ state are exchanged in the $t$-channel. One the motivations for the study of the $\eta_c$ production is that the meson mass provides a hard scale that makes a perturbative calculation possible. Unfortunately, 
 the $ep$ cross section for this process was  too small to be observed at HERA and a future separation of the exclusive $\eta_c$ photoproduction in photon - induced interactions at the LHC is expected to be a very hard task \cite{vicbru_etac, kleinetac}.   

\begin{widetext}

\begin{figure}[t]
\begin{tabular}{ccc}
\includegraphics[width=0.575\textwidth]{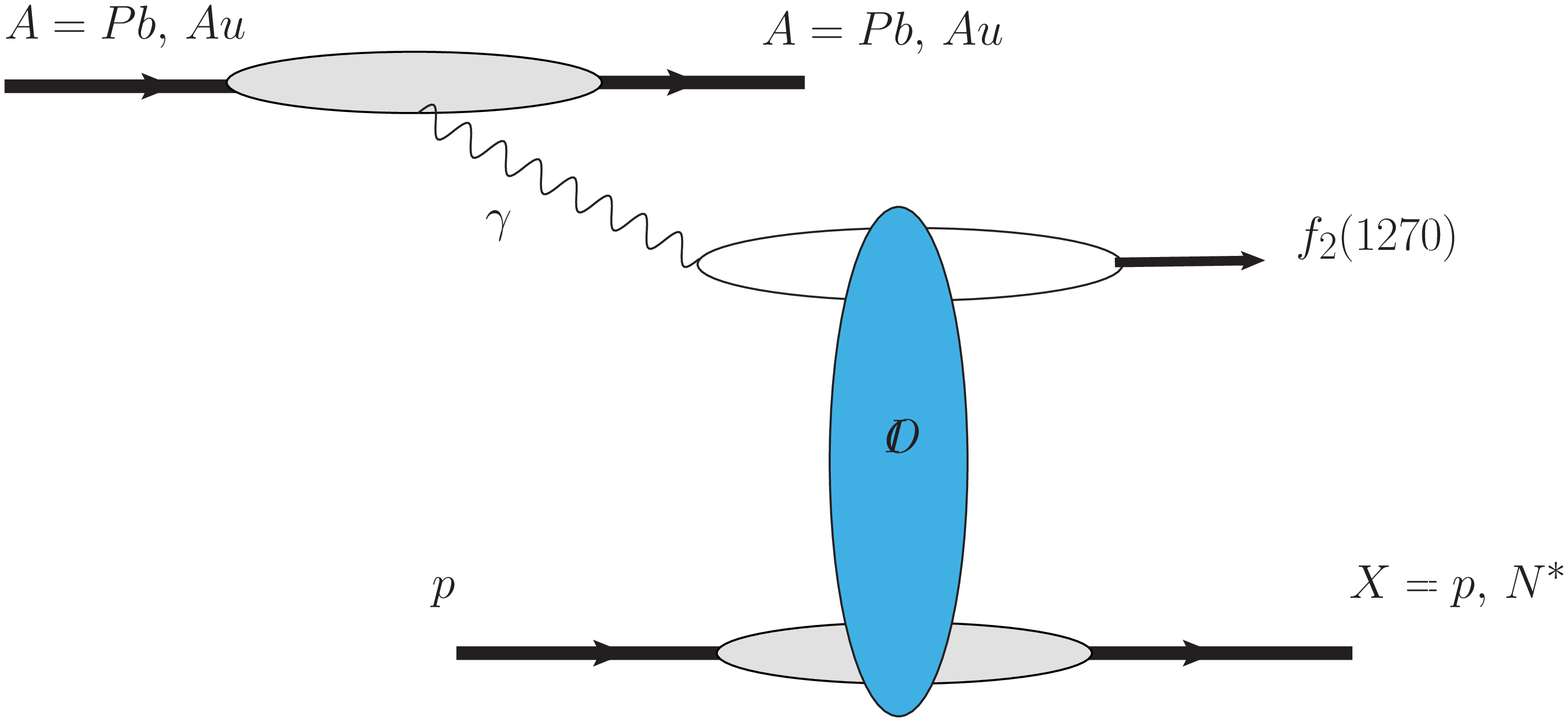} & \,\,\hspace{1cm}&
\includegraphics[width=0.33\textwidth]{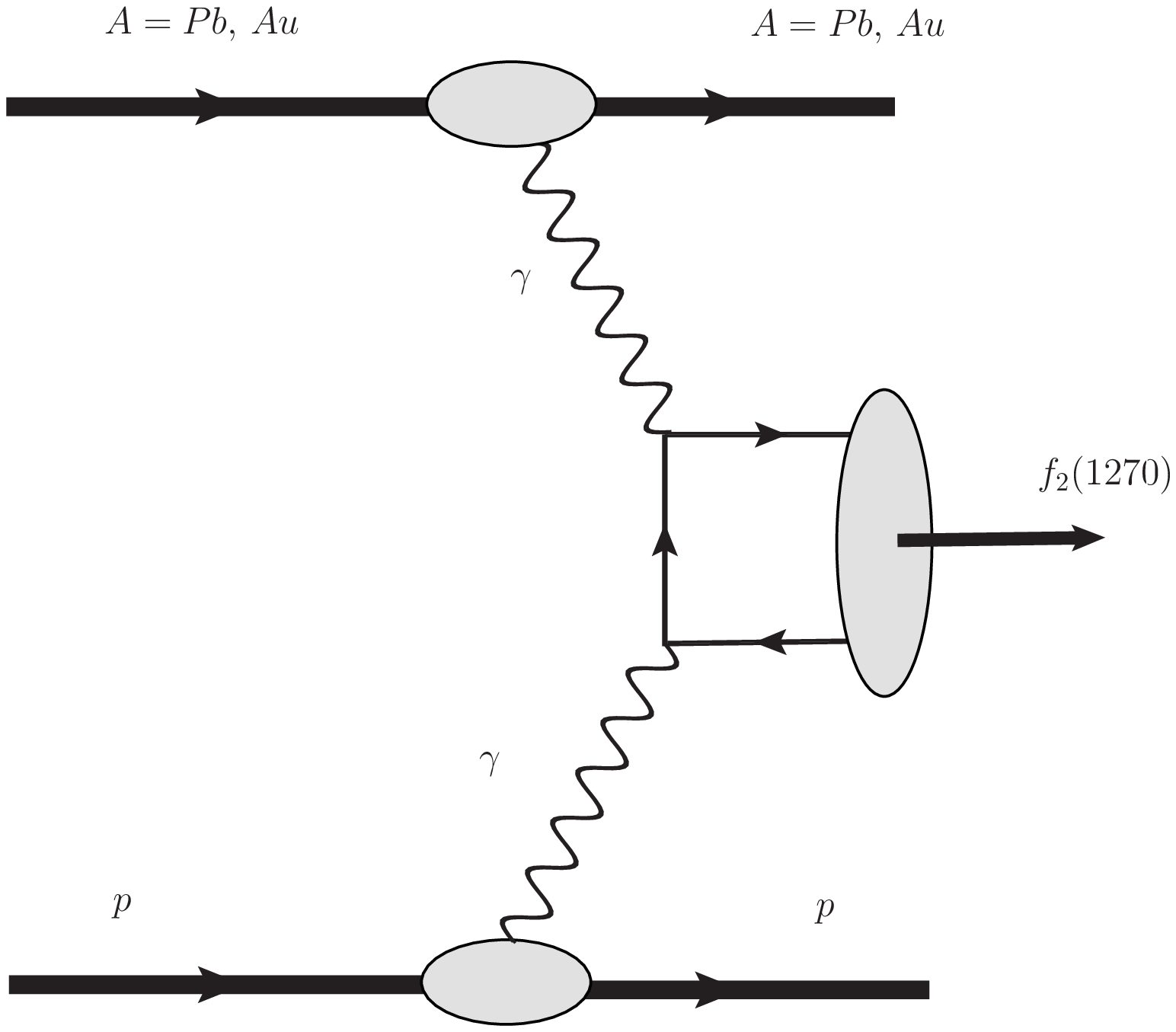} \\
(a) & \,\,& (b)
\end{tabular}
\caption{$f_2(1270)$ prodution in (a) photon -- Odderon and (b) photon -- photon interactions.}
\label{fig1}
\end{figure}

\end{widetext}

In this paper we propose the search of the Odderon  in the diffractive $f_2(1270)$ photoproduction in $pA$ collisions at RHIC and LHC. Such process is represented in Fig. \ref{fig1} (a). Our motivation for the study of this process is twofold. First, due to the smaller $f_2$ mass, the corresponding cross section is expected to be larger than the $\eta_c$ one. Second,  in $pA$ collisions, due to $Z^2$ dependence of the nuclear photon flux,  the $f_2$ production by photon - Odderon interactions  is expected to be dominant in comparison to that associated to Pomeron - Pomeron reactions. The main background in this case becomes the $f_2(1270)$ production by photon -- photon interactions, as represented in Fig. \ref{fig1} (b), which also will be estimated in what follows.
The main shortcoming in the study of the diffractive $f_2$ photoproduction is associated to the fact that a perturbative approach is, in principle, not justified to calculate the cross section. As a consequence, we should to assume a non -- perturbative model to describe the process. In our analysis we will use the approach proposed in Refs. \cite{nac,nac_odd}, which is based on functional integral techniques and the model of the stochastic vacuum to treat QCD in the non-perturbative region \cite{svm}. In this model the Odderon exchange is calculated from the functional integral of two lightlike Wegner -- Wilson loops, with the resulting cross section being energy independent. Such energy behaviour  is also expected in perturbative QCD as demonstrated by  Bartels, Lipatov and Vacca (BLV) \cite{blv}, which  have found a solution for the BKP equation with intercept $\alpha_{\odd}$ exactly equal to one. Our motivation to use the stochastic vacuum model is directly associated to the fact that this model is able to provide an unified description of a large set of hadronic reactions,  dominated by soft interactions and Pomeron exchange, with a satisfactory agreement with the experimental data (See e.g. Ref. \cite{Shoshi:2002in}). As we will demonstrate in what follows, our results indicate that the diffractive $f_2$ photoproduction in $pA$  collisions at RHIC and LHC is dominant at central rapidities, which implies that a future experimental analysis of this process can be useful to probe the Odderon.

Initially, let's present a brief review of the description of photon -- induced interactions in hadronic collisions at large impact parameter ($b > R_{h_1} + R_{h_2}$) and at ultra relativistic energies. In this regime the  electromagnetic interaction is dominant \cite{upc}.
The photon stemming from the electromagnetic field
of one of the two colliding hadrons can interact with one photon of
the other hadron (two-photon process) or can interact directly with the other hadron (photon-hadron
process). The total
cross section for a given process can be factorized in terms of the equivalent flux of photons of the hadron projectile and  the photon-photon or photon-target production cross section \cite{upc}.   In the particular case of proton -- nucleus collisions, the photon -- hadron interactions are dominated by processes where the nucleus is the source of photons, due to the $Z^2$ dependence of the photon flux,  and the proton is the target. The cross section for the  diffractive $f_2(1270)$ photoproduction in a given rapidity range can be expressed as follows
\begin{eqnarray}
\sigma (A p \rightarrow A \otimes f_2(1270) \otimes X) =  \int_{Y_{min}}^{Y_{max}} dY \frac{d\sigma}{dY}\,,
\label{sighh}
\end{eqnarray}
with
\begin{eqnarray}
\frac{d\sigma}{dY} = \int_{b_{min}} d^2\rb \,\omega N_A (\omega,\rb) \,\sigma_{\gamma p \rightarrow f_2(1270) X} (W_{\gamma p}^2) \,\,\,\,,
\label{rapdis}
\end{eqnarray}
where the rapidity $Y$ of the  meson in the final state is determined by the photon energy $\omega$ in the collider frame and by mass $M_{f_2}$ of the  meson [$Y = \ln (2\omega/M_{f_2})$]. Moreover, the symbol $\otimes$ represents a rapidity gap in the final state, $X = p$ or  $N^*$ depending if the proton remains intact or is excited for a $2P$ state in the interaction.  ${d\sigma}/{dY}$ is the rapidity distribution for the photon--proton interaction induced by the nucleus $A$ and $N(\omega,\rb)$ is the equivalent photon flux for a given 
 photon energy $\omega$ and impact parameter $\rb$, which can be expressed in terms of the nuclear form factor $F$ as follows
\begin{widetext}
 \begin{equation}
N_A(\omega,b) = \frac{Z^{2}\alpha_{em}}{\pi^2}\frac{1}{b^{2}\omega}
\left[ \int u^{2} J_{1}(u) F\left(\sqrt{\frac{\left( {b\omega}/{\gamma}\right)^{2} + u^{2}}{b^{2}}} \right )
\frac{1}{\left({b\omega}/{\gamma}\right)^{2} + u^{2}} \mbox{d}u\right]^{2}\, .
\label{fluxo}
\end{equation}
\end{widetext}
As in previous studies \cite{vic_odderon1,vic_odderon2, vicbru_etac} we will assume a monopole form factor given by \cite{kluga}
\begin{equation}
F(q) = \frac{\Lambda^{2}}{\Lambda^{2} + q^{2}} \, ,
\label{ff_nuc}
\end{equation}
with $\Lambda = 0.088$ GeV and $b_{min} = R_A + R_p$. We have verified that the results  obtained using a realistic nuclear form factor differ by less than $3\%$, which is expected since we are assuming that $b \ge b_{min}$ \cite{kluga,Azevedo:2019fyz}.   Finally, $W_{\gamma p}^2=2\,\omega \sqrt{s_{\mathrm{NN}}}$  and ${s_{\mathrm{NN}}}$ are  the  c.m.s energy squared of the
photon - proton and nucleus-proton system, respectively. The cross section $\sigma_{\gamma p \rightarrow f_2(1270) X}$ is given by
\begin{eqnarray}
\sigma({\gamma p \rightarrow f_2(1270) X}) = \frac{1}{16 \pi} \int dt \,\, |\langle \Phi_{\gamma f_2}|{{\cal{J}}_{MSV}}| \Phi_{pX} \rangle|^2\,\,,
\label{sig_gp}
\end{eqnarray}
where $\Phi^i_{\gamma f_2}$ and $\Phi_{pX}$  are the impact factors for the $\gamma \rightarrow f_2$ and $p \rightarrow X$ transitions, respectively. Moreover, 
${{\cal{J}}_{MSV}}$ describes the interaction between the $q \bar{q}$ dipole and the proton structure as predicted by the model of stochastic vacuum  (MSV) \cite{nac,nac_odd} (We refer the reader to the original papers for the details).  The quantity $\Phi_{\gamma f_2}$  can be calculated in terms of the photon wave function and the light cone wave function for the tensor meson $f_2$. On the other hand, 
the impact factor $\Phi_{pX}$ describes the coupling of the Odderon to the proton and takes into account  the transition of the proton into a state $X$, which is assumed to be a proton or an excited $N^*$ state. Such impact factor depends on the proton wave function and on the wave function for the 2P resonance, which are non-perturbative quantities and should be modelled. As in Refs. \cite{nac,nac_odd} the proton will be assumed as having a quark -- diquark structure, which implies that 
${{\cal{J}}_{MSV}}$ can be estimated considering a dipole -- dipole interaction mediated by an Odderon exchange. This assumption leads to a reduction in the proton -- Odderon -- proton coupling, which implies that the Odderon contribution for the elastic scattering is negligible at small -- $t$. In addition, the cross section for the  
$\gamma p \rightarrow f_2(1270) p$ is strongly suppressed in comparison to the case where the initial proton is transformed diffractively into an excited negative parity state, as e.g. $N^* = N(1520)$ or $N(1535)$. As a consequence, in what follows we  only will present results for the case where $X = N^*$, i.e. for the reaction $\gamma p \rightarrow f_2(1270) N^*$, with the resulting predictions being the sum of the $N(1520)$ and $N(1535)$ contributions. In our calculations we assume the parameters of the model to be the same as those used in Ref. \cite{nac_odd}, which were fixed at an energy of $W_{\gamma p} = 20$ GeV.

The Odderon and the photon exchanged in Fig. \ref{fig1} (a) are colorless objects, which lead to the formation of two rapidity gaps in the final state, i.e. the outgoing particles ($A$, $f_2$ and $X$) are separated by a large region in rapidity in which there is no additional hadronic activity observed in the detector. 
Two rapidity gaps in the final state are also  generated in the $f_2$ production by photon - photon interactions. Therefore it is important to estimate the background associated to this process.
The cross section for the $f_2(1270)$ production by $\gamma \gamma$ interactions [See Fig. \ref{fig1} (b)]  can be expressed by
  \cite{upc}
  \begin{eqnarray}
\sigma \left( A p \rightarrow A \otimes f_2(1270) \otimes p \right)   
&=& \int \hat{\sigma}\left(\gamma \gamma \rightarrow f_2 ; 
W_{\gamma \gamma
} \right )  N_A\left(\omega_{1},{\mathbf b_{1}}  \right )
 N_p\left(\omega_{2},{\mathbf b_{2}}  \right ) S^2_{abs}({\mathbf b})  
\frac{W_{\gamma \gamma
}}{2} \mbox{d}^{2} {\mathbf b_{1}}
\mbox{d}^{2} {\mathbf b_{2}} 
\mbox{d}W_{\gamma \gamma
} 
\mbox{d}Y \,\,\, ,
\label{sec_gg}
\end{eqnarray}
where $W_{\gamma \gamma
} = \sqrt{4 \omega_1 \omega_2}$ is the invariant mass of the $\gamma \gamma$ system, $Y$ is the rapidity of the meson in the final state and  $\sigma_{\gamma \gamma \rightarrow f_2}(\omega_{1},\omega_{2})$ 
is the cross section for the $f_2$ production  by the interaction of two real photons with energies $\omega_1$ and $\omega_2$. The factor $S^2_{abs}({\mathbf b})$ is the absorption factor, given in what follows by \cite{BF90}
\begin{eqnarray}
S^2_{abs}({\mathbf b}) = \Theta\left(
\left|{\mathbf b}\right| - R_{A} - R_{p}
 \right )  = 
\Theta\left(
\left|{\mathbf b_{1}} - {\mathbf b_{2}}  \right| - R_{A} - R_{p}
 \right )  \,\,,
\label{abs}
\end{eqnarray}
where $R_{h}$ is the radius of the hadron $h$ ($h = A,\,p$). 
The presence of this factor in Eq. (\ref{sec_gg})  excludes the overlap between the colliding hadrons and allows to take into account only ultraperipheral collisions.
Moreover, the cross section for the production of  the $f_2$ 
state due to the two-photon fusion can be written in terms of the two-photon decay width of  the corresponding state as  follows \cite{Low}
\begin{eqnarray}
 \sigma_{\gamma \gamma \rightarrow f_2}(\omega_{1},\omega_{2}) = 
8\pi^{2} (2J+1) \frac{\Gamma_{f_2 \rightarrow \gamma \gamma}}{M_{f_2}} 
\delta(4\omega_{1}\omega_{2} - M_{f_2}^{2}) \, ,
\label{Low_cs}
\end{eqnarray}
where the decay width $\Gamma_{f_2 \rightarrow \gamma \gamma}$ is taken from experiment  and $M_{f_2}$ and $J$ are, respectively, the mass and spin of the  $f_2$  state. In our calculations we will assume the values of $\Gamma_{f_2 \rightarrow \gamma \gamma}$
and $M_{f_2}$ as given in Ref. \cite{pdg}.

\begin{table}[t]
\begin{center}
\begin{tabular}{||c|c|c|c||}
\hline 
 & {\bf ALICE/CMS ($-2.0 \le Y \le +2.0$)} & {\bf LHCb ($2.0 \le Y \le 4.6$)} & {\bf RHIC ($-1.0 \le Y \le +1.0$)} \\
\hline
\hline 
$\gamma \odd$ & 8.37  & 0.85 & 0.34 \tabularnewline
\hline 
$\gamma \gamma$ & 3.34 & 5.32 & 0.21 \tabularnewline
\hline
\end{tabular}
\caption{Cross sections in $\mu  b$ for the $f_2(1270)$ production at different rapidity ranges in photon -- Odderon ($\gamma \odd$) and photon -- photon ($\gamma \gamma$) interactions  in $pPb$ collisions at LHC ($\sqrt{s} = 8.1$ TeV) and $pAu$ collisions at RHIC ($\sqrt{s} = 0.2$ TeV).  }
\label{I}
\end{center}
\end{table}

In Table \ref{I} we present our predictions for the $f_2(1270)$ production at different rapidity ranges in photon -- Odderon ($\gamma \odd$) and photon -- photon ($\gamma \gamma$) interactions. We consider  $pPb$ collisions at LHC ($\sqrt{s} = 8.1$ TeV) and $pAu$ collisions at RHIC ($\sqrt{s} = 0.2$ TeV). Moreover, we assume
the rapidity ranges probed by the ALICE/CMS and LHCb detectors at the LHC  as well as the central rapidity range probed in hadronic collisions at RHIC. We obtain cross sections of the order of $\mu  b$, with the $f_2$ production by $\gamma \odd$ interactions being larger than the $\gamma \gamma$ contribution  at mid - rapidities. In contrast, the $\gamma \odd$  contribution is strongly suppressed at forward rapidities due to the energy independence of the $\gamma p \rightarrow f_2 N^*$ cross section and  the decrease of the photon flux at large photon energies.
In comparison to the results for the exclusive $\eta_c$ photoproduction in $pPb$ collisions presented in Ref. \cite{vicbru_etac}, our predictions for $f_2$ are larger by  a factor $\approx 300$. Another advantage of the $f_2$  is that it dominantly decay in a pair of pions, with a branching fraction of $ 84.2 \%$. In contrast, the branching fractions for the  decay of the $\eta_c$ into stable hadrons are smaller than $17.4 \%$. Considering that the CMS integrated luminosity for $pPb$ collisions at $8.1$ TeV in 2016 was  $\approx$ 180/nb, we predict that number of diffractive events will be larger than $1.5 \times 10^6$.  Such results indicate that the experimental analysis of the diffractive $f_2$ photoproduction is, in principle, feasible and can be considered ideal to probe the Odderon.

Some comments are in order. In this paper we have assumed the model of the stochastic vacuum to describe the odderon exchange, which in the case of the $f_2$ production is expected to be dominated by non -- perturbative physics. As a consequence, our predictions should be considered model dependent. Moreover, as in Ref. \cite{nac_odd} we estimate that the  uncertainty in our predictions to be a factor $\approx 2$, which is directly associated to the treatment of the $\Phi_{pN^*}$ impact factor. Certainly the calculation of diffractive $f_2$ photoproduction assuming a different approach is a subject that deserves to be considered. We expect that the results presented here motivate a future study.  An alternative is to use the approach proposed in Refs. \cite{Ewerz:2013kda,Lebiedowicz:2013ika}, which sucessfully describes  a large set of soft processes. Taking into account the uncertainty in our results, we have that the $\gamma \odd$ and $\gamma \gamma$ predictions become similar. In principle such contributions can be separated assuming a lower cutoff in the transverse momentum of the $f_2$ state, since $\gamma \odd$ interactions are expected to generate this state with a larger $p_T$ in comparison to those produced by $\gamma \gamma$ fusion. It is important to emphasize that the theoretical uncertainty in the cross section for $f_2$ production by 
$\gamma \gamma$ interactions is very small. Consequently, if a larger value is experimentally observed, such enhancement can be directly associated to the $\gamma\odd$ contribution and can be considered an evidence of the Odderon.

Finally, let's  summarize our main results and conclusions.
One the main open questions in the field of strong interaction physics is the description of the diffractive processes. In particular, the existence of the Odderon, which is an unambiguous prediction of QCD,  is still not confirmed in the experiment. 
Certainly, a probe of its existence (or not) will improve our undertanding about the diffractive interactions in QCD. In this paper we have proposed to probe the Odderon in photon -- induced interaction at $pA$ collisions. We  have studied the diffractive $f_2(1270$ photoproduction in which the diffractive interaction is described by an Odderon exchange, with the Pomeron one being suppressed by the quantum numbers of the final state. Therefore, the observation of such processes would clearly indicate the existence of the Odderon. We predict  total cross sections of order of $\mu b$ for the $f_2$ production at midrapidities in $pA$ collisions which makes, in principle, the experimental analysis of this process feasible at LHC and RHIC.

\section*{Acknowledgements}
VPG acknowledge useful discussions with Spencer Klein, Valery Khoze,  Bruno Moreira, Daniel Tapia - Takaki, Christophe Royon and Antoni Szczurek.  The author is grateful to the members of the Department of Physics and Astronomy of the University of Kansas by the warm hospitality during the initial phase of this study.
 This work was  partially financed by the Brazilian funding
agencies CNPq,  FAPERGS and INCT-FNA (process number 
464898/2014-5).



\end{document}